\documentclass[preprint,showpacs,amsmath,amssymb]{revtex4}

    \usepackage{graphicx}
    \usepackage{dcolumn}
    \usepackage{bm}

    \newcommand{\corr}[1]{{{#1}}}
    \newcommand{\VEC}[1]{\mbox{\boldmath${#1}$}}

    \newcommand{\IgnoreThis}[1]{#1}

\begin{document}

\title{
Optical Lenses for Atomic Beams}

\author{Ole Steuernagel}

\address{School of Physics, Astronomy and Mathematics, University of
Hertfordshire, Hatfield, AL10 9AB, UK }
\email{O.Steuernagel@herts.ac.uk}

\date{\today}

\begin{abstract}
Superpositions of paraxial laser beam modes to generate atom-optical
lenses based on the optical dipole force are investigated
theoretically. Thin, wide, parabolic, cylindrical and circular atom
lenses with numerical apertures much greater than those reported in
the literature to date can be synthesized. This superposition
approach promises to make high quality atom beam imaging and
nano-deposition feasible.
\end{abstract}




\maketitle\IgnoreThis{\section{Introduction}}

The field of atom-optics offers considerable potential in applied and
fundamental physics, both for atom beam litho\-graphy (to
create nano-structures)~\cite{Meschede_Metcalf_03,ItoNAT00} and for
atom beam microscopy~\cite{Doak_PRL99,Ward_07}. Here, the use of the
optical dipole force using far detuned laser light for the
manipulation of neutral atoms is considered. In this regime it yields
a conservative potential for the manipulation of atoms that is
proportional to the laser light
intensity~\cite{Metcalf.book,Meschede_Metcalf_03}.

Already in 1978 Ashkin's group demonstrated neutral atom beam
focusing using the optical dipole force~\cite{Bjorkholm_Ashkin78}.
Many techniques to focus atomic beams have been tried since:
mirrors~\cite{HolstNAT97_Mirror,Barredo_AM_08}, transmission
gratings~\cite{Keith_PRL88,Carnal_Mlynek_PRL91}, holographic
reflection-gratings~\cite{Oberst_Shimizu_PRL05}, electro-static
lenses~\cite{Noh_Shimizu_EstatLens_PRA00}, magnetic
lenses~\cite{Drndic_APL98,Kaenders_PRA96,Meschede_Metcalf_03,Marechal_APB08,Smith_JOPA08}
or magnetic mirrors~\cite{Hinds99},
nano-apertures~\cite{Balykin_OPN05,Muetzel_APB05,Ito_JOPA06}, and
optical
setups~\cite{Helseth_PRA02,Gangat_PRA05,Balykin_OPN05,Ito_JOPA06,
  Bjorkholm_Ashkin78,Gallatin_1991OSAJB,McClelland_1991OSAJB,Timp_PRL92,Natarajan96,
  Johnson_Prentiss_SCI98,Muetzel_PRL02,Meschede_Metcalf_03,Dubetsky_ConicalLaser_PRA98,
  Muetzel_APB03,Klimov_JMO95} relying on the optical dipole
force~\cite{Metcalf.book,Meschede_Metcalf_03}.

Amongst optical dipole force approaches there are schemes based on
pulsed laser configurations~\cite{Helseth_PRA02,Gangat_PRA05}, light
confined by nano-apertures~\cite{Balykin_OPN05}, single-mode hollow
beams~\cite{Bjorkholm_Ashkin78,Gallatin_1991OSAJB,McClelland_1991OSAJB}
or standing wave setups that yield tightly spaced ridges of the
atomic deposition
patterns~\cite{Timp_PRL92,Natarajan96,Johnson_Prentiss_SCI98,Muetzel_PRL02,Meschede_Metcalf_03}.
Standing wave pattern approaches can also yield other deposition
patterns~\cite{Dubetsky_ConicalLaser_PRA98,Muetzel_APB03}, but
because of the high spatial frequencies involved, smooth profiles
such as those desired for aberration-free atom-lenses wider than 200
nm cannot be synthesized with this
approach~\cite{Klimov_JMO95,Natarajan96,Meschede_Metcalf_03,Muetzel_APB03}.

In the case of standing wave
setups~\cite{Meschede_Metcalf_03,Muetzel_APB03} spherical
aberrations give rise to pronounced pedestals, filling the gaps
between patterned
areas~\cite{Timp_PRL92,Meschede_Metcalf_03,Muetzel_APB03}. This
makes it impossible to lay down separate nano-wires. A pulsed
approach should reduce the pedestal problem~\cite{Barberis_PRA03}
but remains constrained by the short spatial wavelengths typical for
standing wave approaches~\cite{Ritt_Weitz_PRA06}. A related
approach~\cite{Johnson_Prentiss_SCI98}, that suffers less from
pedestal problems, uses atomic de-excitation processes creating an
effective transmission mask for excited noble-gas atoms to etch
structures. Unfortunately, it appears to be unsuitable for direct
deposition of metal atoms (they tend to stick to the deposition area
regardless of their internal state). Its inherent filtering reduces
atomic deposition rates and, more importantly, it does not redirect
the center of mass of atomic motion and thus cannot be used for
traditional imaging of atomic beams.

Similar problems occur in the application of single-mode hollow laser
beams~\cite{Gallatin_1991OSAJB,McClelland_1991OSAJB} as optical
imaging elements. Their waist is potentially wide, but their
elongation leads to thick lenses with small numerical apertures: for
realistic set\-ups a diameter of the transverse parabolic part of the
potential of less than 200 nm arises in conjunction with focal lengths
in the micrometer
range~\cite{Gallatin_1991OSAJB,McClelland_1991OSAJB,Balykin_OPN05}
yielding unsatisfactorily small numerical apertures for atomic
focussing. This implies that one would have to start out with already
well focussed atomic beams and, yet, the resulting atomic point-spread
function remains unsuitably wide. None of the approaches mentioned so
far has been adopted as a solution for the problem of imaging of
atomic beams in atomic microscopy~\cite{Doak_PRL99,Ward_07} or direct
atom-deposition litho\-graphy~\cite{Meschede_Metcalf_03,ItoNAT00}: a
viable atom-optical lens still needs to be found.

\begin{figure}[h]
\centering\includegraphics[width=7cm]{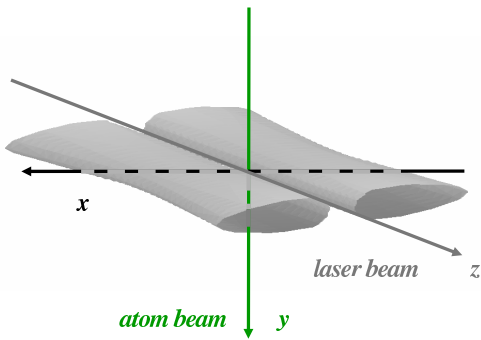} \caption{(Color
online) Arrangement of beams.} \label{fig_geometry}
\end{figure}

\begin{figure}[t]
\centering
\includegraphics[width=3.3in,height=2.5in]{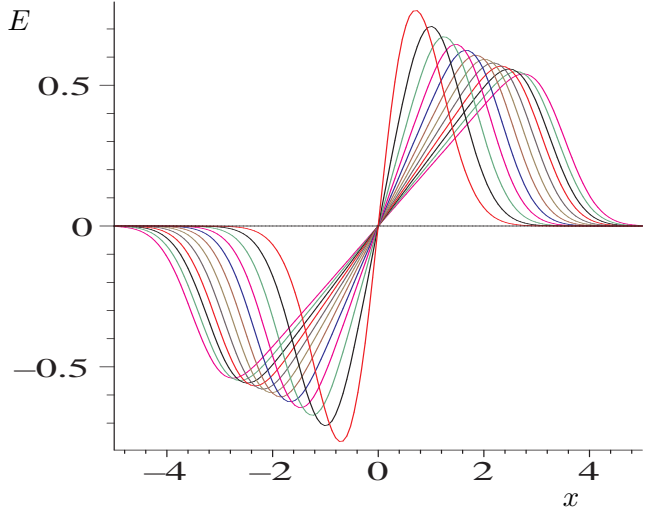}
\put(-240,170){\rotatebox{0}{\mbox{$E$}}} \put(-30,-10){$x$}
\caption{(Color online) Electric field profile,~$E_{2J+1}(x,0,0)$,
at focal cross-section of Hermite-Gauss beams comprising
superpositions of up to 23rd order odd modes (i.e. $2J+1 =
1,3,\ldots,23$; $x$-axis in units of beam waist~$w_{0x}$, total
cross-sectional beam power normalized to unity, Rayleigh lengths
kept constant, ${\epsilon_0 \omega_L^2/2}$ set to unity).}
\label{fig_Efield}
\end{figure}

Here, it is shown that only the superposition of many laser
modes~\cite{OleJMO05,Ole_2005AmJPh,Maurer_RitschMarte_NJP07} will
allow us to generate wide atom-optical lenses based on the optical
dipole force. We will find that widening the beams' waists is not a
solution if atomic lenses with large numerical apertures are desired,
because prohibitive increases in laser power are necessary.  The idea
of this paper is to superpose several {odd} Hermite-Gaussian
TEM${}_{mn}$-modes,
$\psi_{m,n}$~\cite{Siegman.book,Haus.book,Pampaloni_2004}, such that
all non-linear terms in the dependence of the electric field on the
(transverse) $x$-direction are optimally suppressed, see
Fig.~\ref{fig_geometry}. This generates an electric field profile that
varies linearly across a large part of the laser beam's cross section,
see Fig.~\ref{fig_Efield}, and yields the desired parabolic laser
intensity profile to generate an aberration-free atom-optical lens.

After an introduction of the underlying idea in
Section~\ref{S_odd_Modes}, its possible implementation using
Hermite-Gaussian modes to generate cylindrical lenses is described
in Section~\ref{S_CylindricalLenses}.
Section~\ref{S_SphericalLenses} generalizes this approach to a
crossed beam configuration that yields thin spherical lenses. We
conclude in Section~\ref{S_Conclusion}.

{\section{Superpositions of odd modes}\label{S_odd_Modes}}

We now consider cylindrical atom-lenses with a parabolic modulation
in the $x$-direction, see Fig.~\ref{fig_geometry}; most of what
follows can be translated into the scenario of circular lenses for
which atomic beams co-propagate with the focussing laser
beams~\cite{Bjorkholm_Ashkin78,Gallatin_1991OSAJB} on their optical
axis -- instead of crossing through it. Such circular lenses would
require the use of Laguerre-Gaussian instead of Hermite-Gaussian
modes~\cite{Gallatin_1991OSAJB} but they have the disadvantage of
yielding either tiny lenses (in the case of strongly focussed laser
beams) or thick lenses (for less focussed laser
beams)~\cite{Gallatin_1991OSAJB}. We therefore do not investigate
setups with laser beams co-propagating with the atomic beam here;
instead, we will show in Section~\ref{S_SphericalLenses} how to
create a thin spherical lens using a combination of two orthogonally
crossed multimode Hermite-Gaussian laser beams.

\IgnoreThis{\subsection{Hermite-Gaussian modes}\label{ss_HG-modes}}

Let us consider
modes,~$\psi_{m,n}$~\cite{Siegman.book,Haus.book,Pampaloni_2004},
with transverse beam coordinates $x$ and $y$ propagating in the
$z$-direction. The Rayleigh lengths $z_{R_x}$ and $z_{R_y}$
associated with the two transverse coordinates, $x$ and $y$, can be
different from each other (focussed by different cylindrical lenses,
say). In this case two different associated beam waist radii, $w_{0
x}$ and $w_{0 y}$, and Gouy-phases,~$\phi_{x}(z)$ and~$\phi_{y}(z)$
arise. In the paraxial approximation the normalized modes have the
form
\begin{eqnarray}
\psi_{m,n}({\VEC r})  = &
\sqrt{\frac{\sqrt{2}}{w_x(z)}} \; \;
\varphi_m\left(\frac{\sqrt{2}\,x}{w_x(z)}\right)\; & \exp\left(\frac{i
k_L}{2 } \frac{x^2}{R_x(z)}\right) \; \exp\left(-i (m+\frac{1}{2})
\phi_x(z)\right)
\nonumber \\
 \times & \sqrt{\frac{\sqrt{2}}{w_y(z)}} \; \;
\varphi_n\left(\frac{\sqrt{2}\,y}{w_y(z)}\right)\; & \exp\left(\frac{i
k_L}{2 } \frac{y^2}{R_y(z)}\right) \; \exp\left(-i (n+\frac{1}{2})
\phi_y(z)\right)
\;. \label{EQ_HG.TEM}
\end{eqnarray}
Here, $\VEC r = (x,y,z)$ is the position vector, $\omega_L$ the
frequency of the monochromatic laser, $k_L = \omega_L/c = 2 \pi /
\lambda_L$ its wavenumber, and~$\varphi_m(\xi) =
H_m(\xi)\,\exp(-\xi^2/2) / \sqrt{2^m m! \sqrt{\pi} } $,
$(m=0,1,2,\ldots),$ with the Hermite
polynomials~$H_m$~\cite{Siegman.book,Haus.book,Pampaloni_2004}. The
wave front radii $R(z)= (z^2 + z_R^2)/z$, the beam radii
$w(z)=w_0 \sqrt{ 1+z^2/z_R^2}$ with $w_0=\sqrt{\lambda_L z_R /
\pi}$, and the longitudinal Gouy-phase
shifts~\cite{Siegman.book,Haus.book,Pampaloni_2004},
$\phi(z)=\arctan(z/z_R),$ are all parameterized by the beams'
Rayleigh lengths $z_R$; strictly speaking by $z_{R_x}$ and
$z_{R_y}$, respectively.

In a configuration, such as that displayed in Fig.~\ref{fig_geometry},
one can generate~\cite{OleJMO05,Maurer_RitschMarte_NJP07} a wide
cylindrical atom-lens using a laser beam with an electric field
composed of a suitable combination of odd modes
\begin{equation}
{\Psi}_{2J+1}(\VEC r) = \sum_{j=0}^{J} c_{2j+1} \,
\psi_{2j+1,0}(\VEC r) \; .
 \label{eq_OddModes}
\end{equation}
Here, the beam is modulated in the $x$-direction whereas for the
$y$-direction the purely Gaussian lowest order mode $\varphi_0$ is
employed. Note that this allows us to make the lens `thin' in the
$y$-direction. With increasing cutoff,~$J$, the superposition
pattern becomes increasingly dephased due to the action of Gouy's
phase~\cite{Ole_2005AmJPh}, this will be further investigated in
Section~\ref{S_SphericalLenses}.

Following reference~\cite{Haus.book} the modes in
Equation~(\ref{EQ_HG.TEM}) yield an electric field which is
polarized in the $y$-direction with a small contribution in the
$z$-direction due to the tilt of wave fronts off the beam axis
($\hat{\bf x}, \hat{\bf
  y}, \hat{\bf z}$ are the unit-vectors and $\Re$ stands for
real-part)

\begin{eqnarray}
{\VEC E}_{2J+1}({\VEC r};t) = \Re \{ [ \hat{\bf y} \; \omega_L \; \Psi_{2J+1}
+ \hat{\bf z} \; ic \; \frac{\partial \Psi_{2J+1}}{\partial x} ] e^{i(k_L
z - \omega_L t)} \} \; .
\label{true.E.field}
\end{eqnarray}

For beams that are not too tightly focused we neglect the transverse
derivatives. The associated time-averaged light intensity distribution
then has the form~\cite{Haus.book}

\begin{eqnarray}
I_{2J+1}({\VEC r}) = \epsilon_0 \; \left\langle {\VEC
E}_{2J+1}({\VEC r},t)^2\right\rangle \approx \frac{\epsilon_0}{2} \;
\omega_L^2 \; | \Psi_{2J+1}({\VEC r}) |^2 \; .
\label{Intensity}
\end{eqnarray}

{\subsection{Normalization and Intensity
Scaling}\label{SS_Norm_Intensity_singleMode}}

With the normalized modes of Eq.~(\ref{EQ_HG.TEM}) and assuming that
the sum of the coefficients $\sum |c_{2j+1}|^2$ in
Eq.~(\ref{eq_OddModes}) is unity we use the normalization

\begin{eqnarray}
\int_{-\infty}^\infty \int_{-\infty}^\infty dx \, dy \,
|\psi_{2J+1}(x,y,z)|^2 =\frac{2}{\epsilon_0\omega_L^2}
\int_{-\infty}^\infty \int_{-\infty}^\infty dx \, dy \,
I_{2J+1}(x,y,z)|^2 =1 \,. \label{Eq_Intensity_normalization}
\end{eqnarray}

Assuming validity of the Raman-Nath approximation of negligible
transverse motion of the atoms~($ (x,z)= const.$)~\cite{Wallis95}, the
atoms experience the $y$-integrated intensity distribution of the
laser field given by

\begin{eqnarray}
\bar{I}_{2J+1}(x,z) = \int_{-\infty}^\infty dy \,I_{2J+1}({\VEC r})
= \frac{\epsilon_0\omega_L^2}{2} \frac{\sqrt{2}}{{w_x(z)}} \; \;
\left|\sum_{j=0}^{J} \varphi_{2j+1}(\frac{\sqrt{2}\,x}{w_x(z)}) \;
e^{-i (j+\frac{1}{2}) \phi_x(z)}\right|^2 .
\label{eq_Integrated_Intensity}
\end{eqnarray}

We note that this integrated intensity $\bar{I}$ of beams of fixed
total power reduces inversely proportionally to their width $w_{0x}$,
that is, their field amplitudes scale with
$w_{0x}^{-1/2}$. Furthermore the field gradients diminish with
$w_{0x}^{-1}$. This implies that the effective curvature of the
integrated laser light intensity,~$|\nabla \Psi|^2$, responsible for
atomic focussing scales with $w_{0x}^{-3}$. We face an unfavourable
cubic scaling with the beam width if we attempt to expand a laser beam
transversally in order to widen the effective lens without weakening
its refractive power.  Additionally, as we will show below, pure modes
have small useful areas to generate lenses, the combination of these
two factors makes a pure mode approach unfeasible.  It forces us to
employ the mode superpositions studied here.

\IgnoreThis{\subsection{Optical Dipole Force}\label{ss_Atoms}}

We assume that the interaction between atoms and the laser light is
well described by a two-level scheme (excited state `$e$'
and ground state `$g$') in rotating wave
approximation with effective atomic line width~$\Gamma$ and
resonance frequency~$\omega=\omega_e-\omega_g$. This leads to the
expression $ I(\VEC r )\,\Gamma^2/(2 I_S) = \Omega(\VEC r )^2$ for
the Rabi-frequency~$\Omega$ as a function of the ratio of the
local laser intensity~$I(\VEC r )$ and the transition's saturation
intensity $I_S 
=\pi h c \Gamma/(3\lambda^3)$~\cite{Natarajan96,Metcalf.book}. With
sufficiently weak laser intensity~$I$ and sufficiently large
detuning~$\delta_\omega=\omega_L-\omega$ of the laser
frequency~$\omega_L$ from the atomic transition
frequency~$\omega$, the AC-Stark shift gives rise to a conservative
optical dipole potential which, to first order in~$I/I_S$, has the
form~\cite{Metcalf.book,Hope_PRA96}
\begin{eqnarray}
U_\omega \approx \frac{\hbar\, \Gamma^2}{8\, \delta_\omega}
\frac{I(\VEC r)}{I_S} \, . \label{dipole_potential_approx}
\end{eqnarray}
We can determine the atomic de~Broglie wave number~$\kappa$ of atoms
with mass $M$ and initial kinetic energy $K_0 = (\hbar
\kappa_{0})^2/(2M)$ in terms of their kinetic energy~$K$. Disregarding
Doppler detuning, and assuming the validity of the Raman-Nath
approximation ($K_0 \gg U_\omega$), this allows us to calculate the
associated phase shift
\begin{eqnarray}
\Delta \phi(x,z) & \approx & \int dy (\kappa(\VEC r) - \kappa_{0})
 = \frac{\sqrt{2MK_0}}{\hbar} \int dy (
\sqrt{\frac{K}{K_0}} -1)  \label{eq_difference_kappa}
\\
& \approx & \frac{\sqrt{2MK_0}}{\hbar} \;  \; \int  dy \;
(\sqrt{1-\frac{U_\omega}{K_0}}-1) \; \approx  \frac{-\sqrt{2M}
 \; \Gamma^2}{ 16 \sqrt{K_0} I_S \delta_\omega} \;\bar{I}(x,z)
\label{Delta_phi_IntensityInt}
\end{eqnarray}
The dependence of Eq.~(\ref{Delta_phi_IntensityInt}) on the inverse
kinetic energy implies that best performance is achieved for
monochromatic atomic beams; the approximations are in accordance with
the Raman-Nath assumption~\cite{Wallis95}. Work by Drewsen \emph{et
al.}~\cite{DrewsenOC96} showed that, for an atom-lens, chromatic
dispersion can be reduced by tilting the laser beam with respect to
the passing atomic beam, but the focal plane would have to be tilted
as well. Such a tilt, however, elliptically stretches out the atomic
beam's point-spread function.

Aside from spherical aberrations, there are detrimental noise
sources due to spontaneous emission of photons and light
fluctuations. These tend to increase with increasing laser intensity
but can be decreased by increased detuning~\cite{Metcalf.book} or
through the use of more complicated optical level
schemes~\cite{Hope_PRA96}. Further discussion of their influences is
beyond the scope of this paper.

\IgnoreThis{\section{Cylindrical Lenses}\label{S_CylindricalLenses}}

\begin{figure}[t]
\centering
\includegraphics[width=3.3in,height=2.5in]{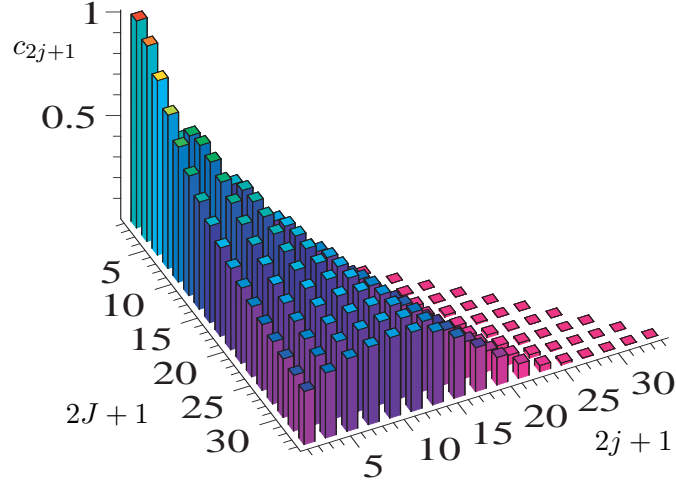}
\put(-250,160){\rotatebox{0}{\mbox{$c_{2j+1}$}}}
\put(-230,20){$2J+1$} \put(-30,10){$2j+1$}
\caption{(Color online) Amplitude coefficients~$c_{2j+1}$ of
Hermite-Gauss superpositions~$\Psi_{2J+1}$ of up to
33${}^{\mbox{rd}}$ order modes ($2J+1 = 1,3,\ldots,33$).}
\label{Fig_coeff_matrix}
\end{figure}

According to eqns.~(\ref{eq_difference_kappa})
and~(\ref{Delta_phi_IntensityInt}) parabolic optical potentials give
rise to parabolic atom-optical phase masks, as is required for
`perfect' atom lenses. In other words, we want the $y$-{integrated}
electric field profile to depend linearly on the $x$-direction, see
Fig.~\ref{fig_Efield}. In order to achieve this we integrate out the
$y$-component, see Eq.~(\ref{eq_Integrated_Intensity}), then
Taylor-expand the field profile and finally choose the coefficients
in Eq.~(\ref{eq_OddModes}) so as to cancel terms non-linear in~$x$.
Using the first $2 J +1$ odd field modes all non-linear terms up to
$(2J+1)^{\mbox{th}}$ order can be cancelled.  The determination of
the coefficients,~$c_{2j+1}$, involves the solution of a linear
equation system and is easily performed. For instance, for the
\corr{third} superposition field,~$\Psi_5$, comprising
Hermite-Gaussian modes $\psi_{1,0},$ $ \psi_{3,0}$ and $\psi_{5,0}$,
the relative strengths of the coefficients are $c_3 = c_1 18
\sqrt{6}/71$ and $c_5 = c_1 2 \sqrt{30}/71 $. For a normalized
superposition the coefficient $c_1$ should be chosen accordingly.
The family of the first \corr{twelve} normalized superpositions
$\{\Psi_{2J+1}, 2J+1=1,3,\ldots,23\}$ is displayed in
Fig.~\ref{fig_Efield}, the associated set of amplitude
coefficients~$c_{2j+1}$ is shown in Fig.~\ref{Fig_coeff_matrix}.

\begin{figure}
\centering
\includegraphics[width=3.4in,height=2.5in]{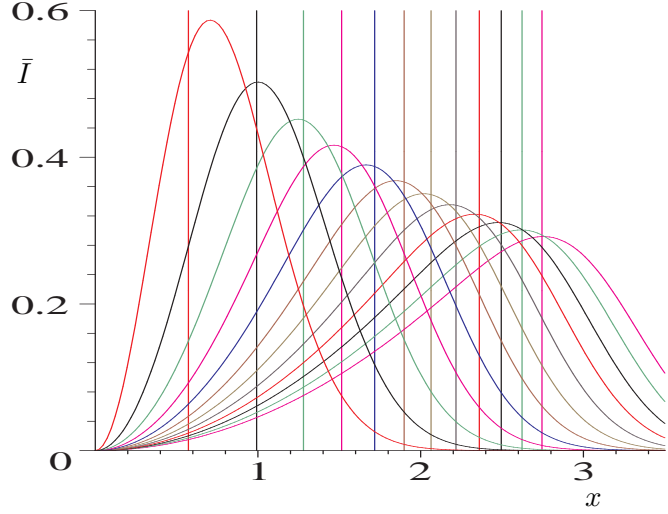}
\put(-245,150){\rotatebox{0}{\mbox{$\bar I$}}} \put(-30,-10){$x$}
\caption{(Color online) Integrated focal intensity profiles~$\bar
I(x,0)$ of Hermite-Gauss superposition beams
  comprising up to 23rd order modes, compare Fig.~\ref{fig_Efield}
  (same units as in Fig.~\ref{fig_Efield}; vertical bars mark location
  of position $0.57\cdot\sqrt{2J+1}\cdot w_{0x}$, confirming harmonic
  oscillator-scaling~\cite{Ole_2005AmJPh}).} \label{fig_I_TurningPoints}
\end{figure}

Mode-superpositions extend the ``useful'' linear part of the field
profile yielding wider parabolic intensity profiles.
Figure~\ref{fig_I_TurningPoints} demonstrates that the useful
parabolic part in the focal intensity profile expands with the
number of superposition modes~$2J+1$ according to the
$\sqrt{2J+1}$-scaling, expected for a harmonic
oscillator~\cite{Ole_2005AmJPh}.
Note, however, that the refractive power of the wider lenses is
reduced (its atom-optical focal length is lengthened), because wider
lenses have reduced transverse field gradients, see
Fig.~\ref{fig_Efield} and discussion following
Eq.~(\ref{eq_Integrated_Intensity}). In order to compensate for this
loss of refractive power, we can increase the transverse field
gradient through either laser beam focussing in the $x$-direction,
or through an increase in laser beam power. Focussing in the
$y$-direction makes no difference since only the {integrated}
intensity $\bar I$ matters. In the next subsection we show how much
the power has to be raised to keep the atomic lenses' refractive
powers equal. Subsequently, in Section~\ref{S_SphericalLenses}, we
will investigate focussing in the $x$-direction; we will see that
Gouy-dephasing constrains this focussing, the lenses must not be
shrunken below a certain limit.

\IgnoreThis{\subsection{Increased Beam Powers Compensates for Lenses
Widening}\label{SS_HG_Mode_Matching}}

\begin{figure}
\centering
\includegraphics[width=3.4in,height=2.5in]{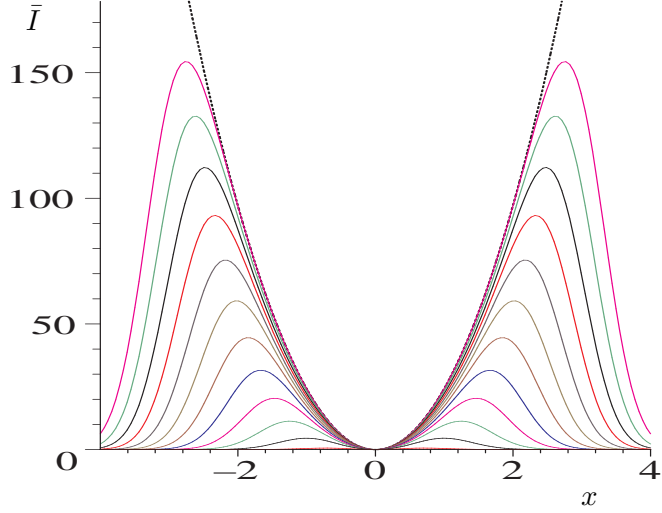}
\put(-240,170){\rotatebox{0}{\mbox{$\bar I$}}}\put(-30,-10){$x$}
\caption{(Color online) Integrated focal intensity profiles~$\bar
I(x,0)$ of
  Hermite-Gauss beams comprising up to 23rd order modes, compare
  Fig.~\ref{fig_Efield} (same units as in
  Fig.~\ref{fig_I_TurningPoints}; total beam power
  adjusted such that all profiles have same curvature at origin
  as the dotted line parabola).} \label{fig_I_equal_Parabola_Power}
\end{figure}

\begin{figure}
\centering
\includegraphics[width=3.4in,height=2.5in]{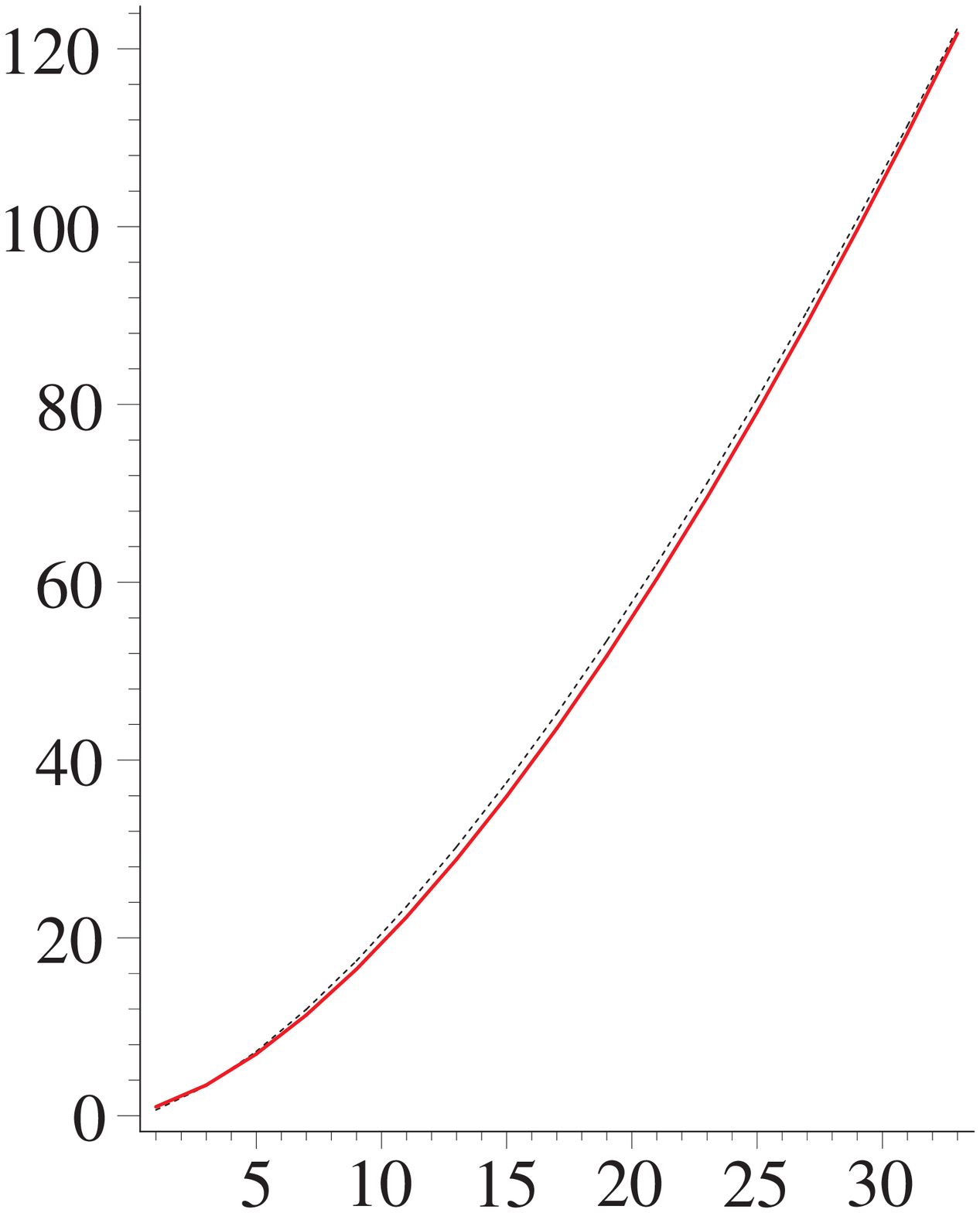}
\put(-260,130){\rotatebox{90}{\mbox{$P_{2J+1}/P_1$}}}\put(-30,-10){$2J+1$}
\caption{(Color online) The increase in total beam power needed to
achieve the power
  compensation described in the text and displayed in
  Fig.~\ref{fig_I_equal_Parabola_Power} as a function of mode number
  (solid red line) scales approximately like $\frac{20}{31}\cdot
  (2J+1)^{3/2}$ (dotted black line).} \label{fig_PowerCompensate}
\end{figure}

If we increase the total beam power $P_{2J+1}$ for wider beam profiles
according to the ratios of the modes' transverse derivatives,
$P_{2J+1} = P_1 |{\partial_x \Psi_1(x,0,0)}/{\partial_x
  \Psi_{2J+1}(x,0,0)}|^2$, the weakened gradient is power-compensated
for by increased laser power. This way all optical potentials give
rise to atom-lenses with equal refractive power, see
Fig.~\ref{fig_I_equal_Parabola_Power}. The necessary beam power
increase to achieve this compensation is sketched in
Fig.~\ref{fig_PowerCompensate}. The power savings due to our multimode
approach are quantified in
Subsection~\ref{SS_Cylindrical_Quality_Power}.

\IgnoreThis{\subsection{Decreased Rayleigh-Lengths Compensate for
Lenses Widening}\label{SS_Rayleigh_Length_Matching}}

\begin{figure}
\centering
\includegraphics[width=3.4in,height=2.5in]{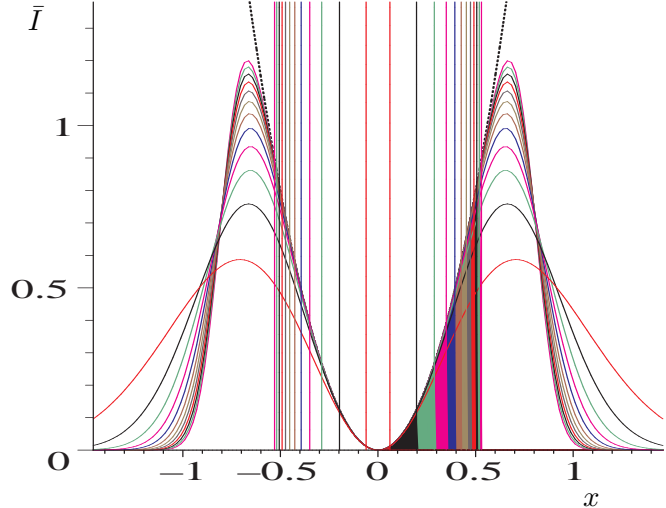}
\put(-240,170){\rotatebox{0}{\mbox{$\bar I$}}} \put(-30,-10){$x$}
\caption{(Color online) Integrated focal intensity profiles~$\bar
I(x,0)$ of
  Hermite-Gauss beams comprising up to 23rd order modes, compare
  Fig.~\ref{fig_Efield}, and their 0.74\%-deviation marks, $d_{2J+1}$,
  which lie at relative positions $ d_{2J+1}/d_1 = 1.00, 3.24,\ldots,
  8.70$ from the origin, compare Table~\ref{table:energy} (same units
  as in Fig.~\ref{fig_Efield}; in contrast to
  Fig.~\ref{fig_I_equal_Parabola_Power} total beam power normalized to
  unity, but Rayleigh lengths~$z_{Rx}$ readjusted such that all
  higher-order superpositions match up with curvature of the first
  mode case~$\Psi_1$, see text).} \label{fig_Iparabola}
\end{figure}

Alternatively to the beam-power increases just discussed, we can
keep the total beam power for all beams equal and shrink the
higher-order superposition-beams' Rayleigh lengths through increased
beam focussing in the $x$-direction. This also allows us to
compensate for the gradient reduction observed in
Fig.~\ref{fig_Efield}. The corresponding laser intensity profiles
are displayed in Fig.~\ref{fig_Iparabola} and lend themselves to an
efficiency analysis of the invested laser power. The vertical bars
in this figure mark the points,~$d_{2J+1}$, where each intensity
curve deviates from the enveloping parabola (dotted line) by 0.74
percent. They delineate the useful areas of the lenses. Beyond a
deviation of 0.74\%, spherical aberrations distort the atomic
point-spread function of an imaged atomic beam too severely. The
filled-in areas under the curves in Fig.~\ref{fig_Iparabola}
represent the laser power fraction contributing to the atom lens in
each case. Higher-order superpositions clearly allow us to use the
laser power much more efficiently. This is quantified in the next
subsection.

\IgnoreThis{\subsection{Lens Quality and Power
Savings}\label{SS_Cylindrical_Quality_Power}}

The 0.74\%-criterion was extracted from the work by Gallatin and
Gould~\cite{Gallatin_1991OSAJB} who considered, for example, the use
of a 0.1 Watt laser detuned by roughly 40,000 linewidths. To achieve
acceptable performance, the effectively useful beam area was found
to be only some $2d_1=$140~nm wide (for a laser beam with a $2
w_{0x}=2\mu$m waist diameter~\cite{Gallatin_1991OSAJB}). In other
words, pure laser modes yield only a small useful window (in order
to fulfill the 0.74\%-deviation criterion only about
$2d_1/(2w_{0x})=$140~\!nm/2$\mu$m$ \, \approx \! 7\%$ of a
cylindrical lens diameter or only the central 0.49 percent area of a
circular lens are useful). Most of the laser power is wasted in the
wings if no suitable superpositions of higher-order modes are
employed.  In our case of a cylindrical lens based on the
Hermite-Gaussian mode~$\psi_{1,0}$, very similarly, approximately
$d_1/w_{0x}=$6\% of the width of the beam is useful, see
Fig.~\ref{fig_Iparabola}. Additionally to the quantification of the
useful area of the lenses, see Table~\ref{table:energy}, this waste
is meaningfully quantified through the determination of the fraction
of power~${\cal E}_{2J+1}$ the laser beam contributes to the
`useful' part of the lens profile. We define it as the ratio of the
laser energy contributing to the area between the deviation
points~$-d_{2J+1}<x<d_{2J+1}$, in terms of the total laser power,
namely

\begin{table}[t]
\caption{Lens Parameters ${d_{2J+1}}$ and ${\cal E}_{2J+1}$, compare
Fig.\ref{fig_Iparabola}}
\centering          
\begin{tabular}{c c c c c c c c c c c c c c c c c c}
\hline\hline                        
$2J+1$ & 1 & 3 & 5 & 7 & 9 & 11 & 13 & 15 & 17 & 19 & 21 & 23
& 25 & 27 & 29 & 31 & 33\\      
\hline                      
${d_{2J+1}}/{d_1}$ & 1.00 & 3.24 & 4.75 & 5.74 & 6.45 & 7.00
& 7.42 & 7.78 & 8.06 & 8.32 & 8.53 & 8.70 & 8.87 & 9.01 & 9.15
& 9.27 & 9.36 \\
${\cal E}_{2J+1}$ [\%] \; \; & 0.048 & 1.6 & 5.1 & 9.1 & 13 & 16
& 20 & 23 & 25 & 28 & 30 & 32 & 33 & 35 & 37 & 38 & 39\\
${{\cal E}_{2J+1}}/{{\cal E}_1}$
& 1 & 34 & 107 & 190 & 269 & 344 & 411 & 472 & 526 & 576 & 620
& 662 & 699 & 735 & 766 & 795 & 825 \\[1ex]
\hline          
\end{tabular}
\label{table:energy}    
\end{table}

\begin{eqnarray}
{\cal E}_{2J+1} = \frac{\int_{-\infty}^{\infty} dy \int_{-d_{2J+1}}^{d_{2J+1}} dx
\; I_{2J+1}(x,y,0)}{\int_{-\infty}^{\infty} dy \int_{-\infty}^{\infty} dx
\; I_{2J+1}(x,y,0)} \, . \label{eq_Rel_power}
\end{eqnarray}

Figure~\ref{fig_Iparabola} and Table~\ref{table:energy} summarize and
quantify our findings.  Specifically, Table~\ref{table:energy}
allows us to compare values for a single-mode atom lens for which
${\cal E}_1 =0.048\%$ with the superposition approach. For example,
compared to mode~$\Psi_1$ the relative power savings in case of
superposition ${\Psi}_{33}$ is 825, this translates into a power
utilization of ${\cal E}_{33}=0.048\% \times 825 = 39 \%$. In
general the details of this behaviour depend on the chosen quality
criterion but the underlying scaling is straightforward to derive.
The useful fraction of the laser beam is proportional to a linear
integral over the intensity and therefore grows with the third power
of the position of the deviation mark ${\cal E}_{2J+1}/{\cal E}_{1}
= (d_{2J+1}/d_1)^3$.

\section{{Spherical Lenses}\label{S_SphericalLenses}}

We now want to investigate the constraints that arise when an
identical copy of the laser beam that travels along the $z$-axis,
see Fig.~\ref{fig_geometry}, is additionally sent along the $x$-axis
such that their crossed configuration leads to the simultaneous
application of two cylindrical lenses giving rise to the application
of a spherical lens to the atomic beam. Either the laser beams are
slightly displaced along the $y$-axis, or they are sufficiently
detuned from each other such that despite their spatial overlap no
harmful interference occurs~\cite{DePue_AntiJitter_PRL99}.

\begin{figure}[ht]
\begin{center}
  \begin{minipage}[b]{0.28\linewidth} 
     \includegraphics[width=1\linewidth,height=1\linewidth]{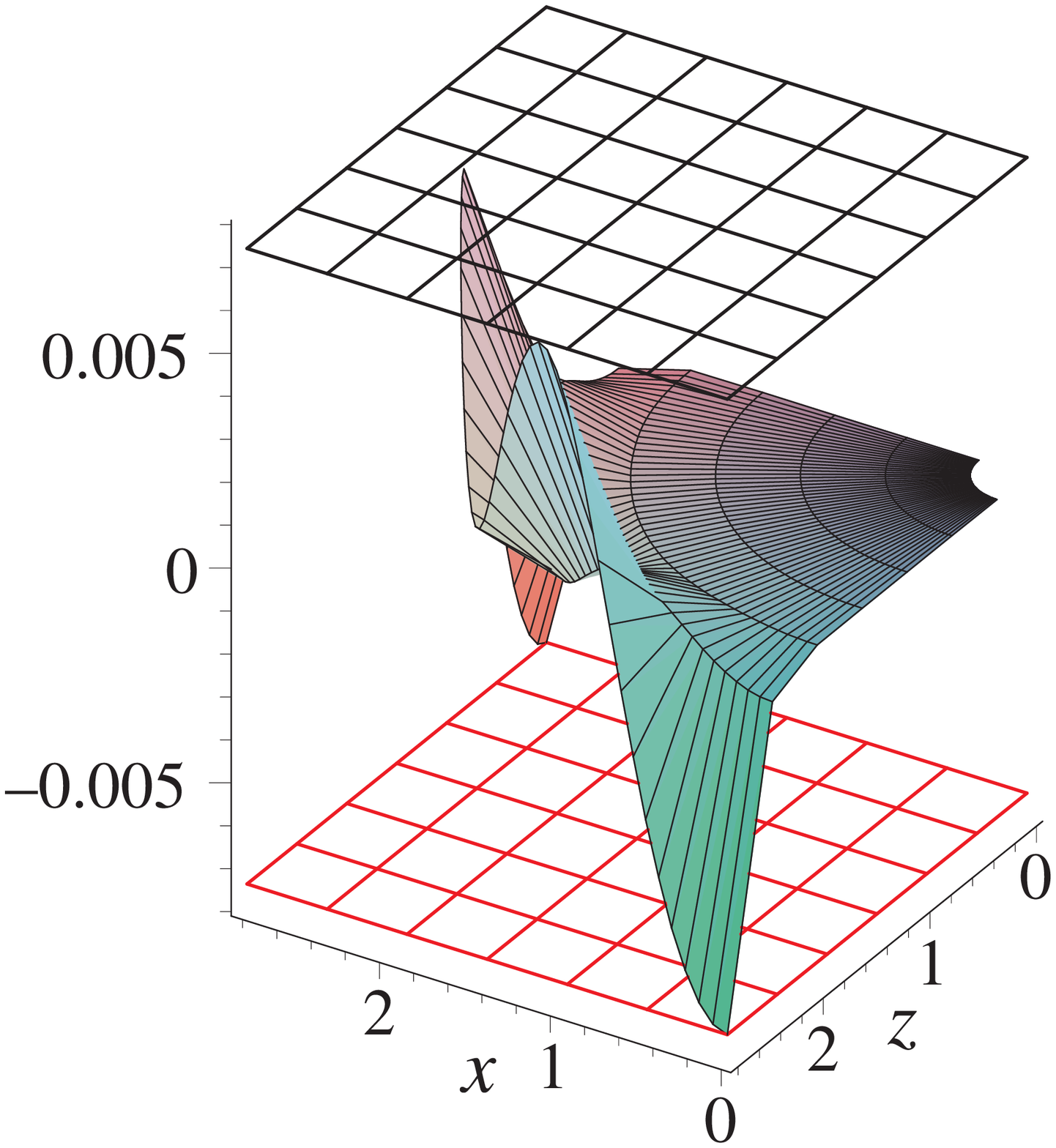}
     \put(-130,110){\rotatebox{0}{\mbox{$\Delta \bar I_{23}$}}} 
  \end{minipage}   
  \hspace{0.045\linewidth}
  \begin{minipage}[b]{0.28\linewidth}
     \includegraphics[width=1\linewidth,height=1\linewidth]{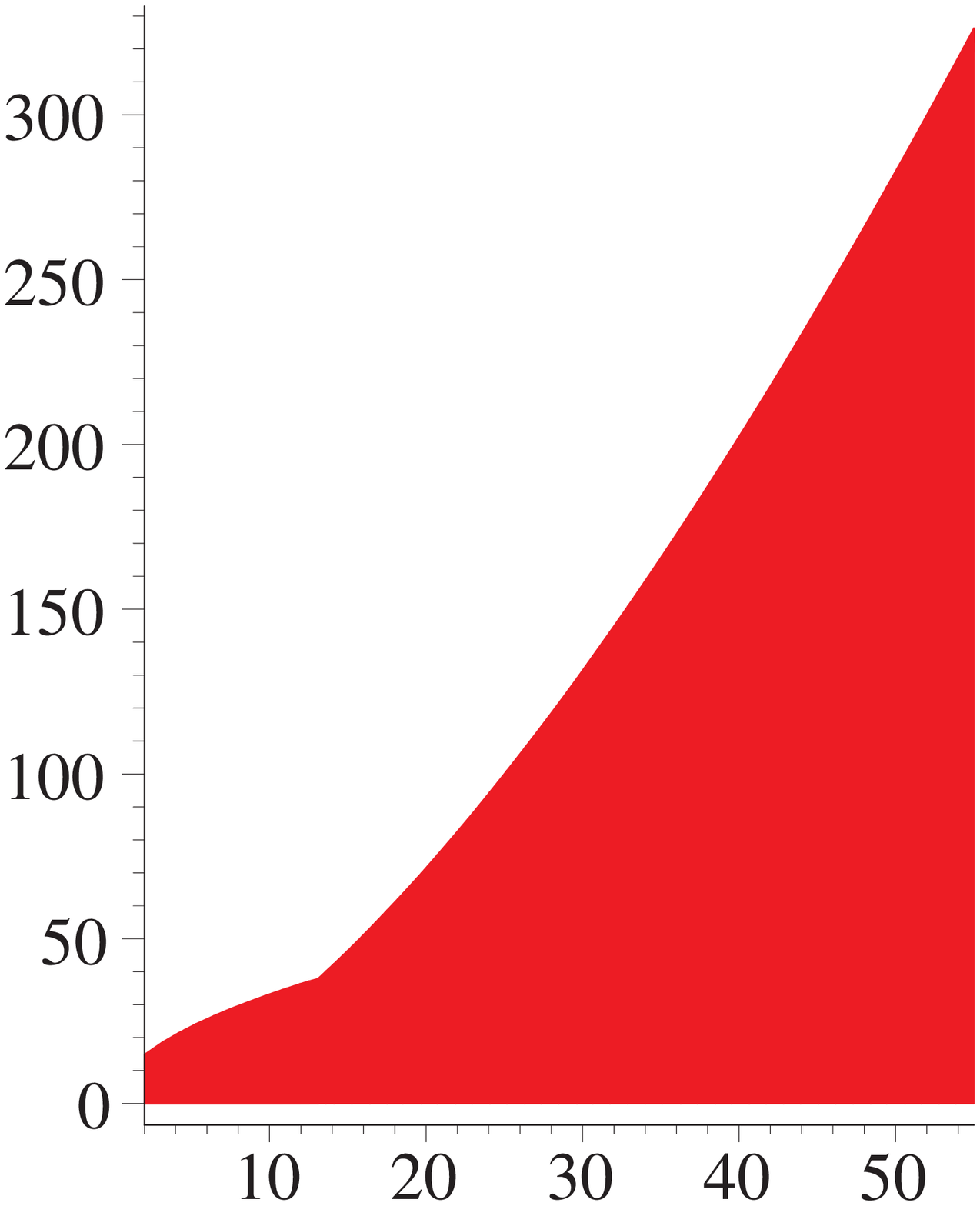}
     \put(-142,50){\rotatebox{90}{\mbox{$z_{\min}(2J+1)/\lambda_L$}}} \put(-30,-10){$2J+1$}
  \end{minipage}
  \hspace{0.045\linewidth}
  \begin{minipage}[b]{0.28\linewidth}
     \includegraphics[width=1\linewidth,height=1\linewidth]{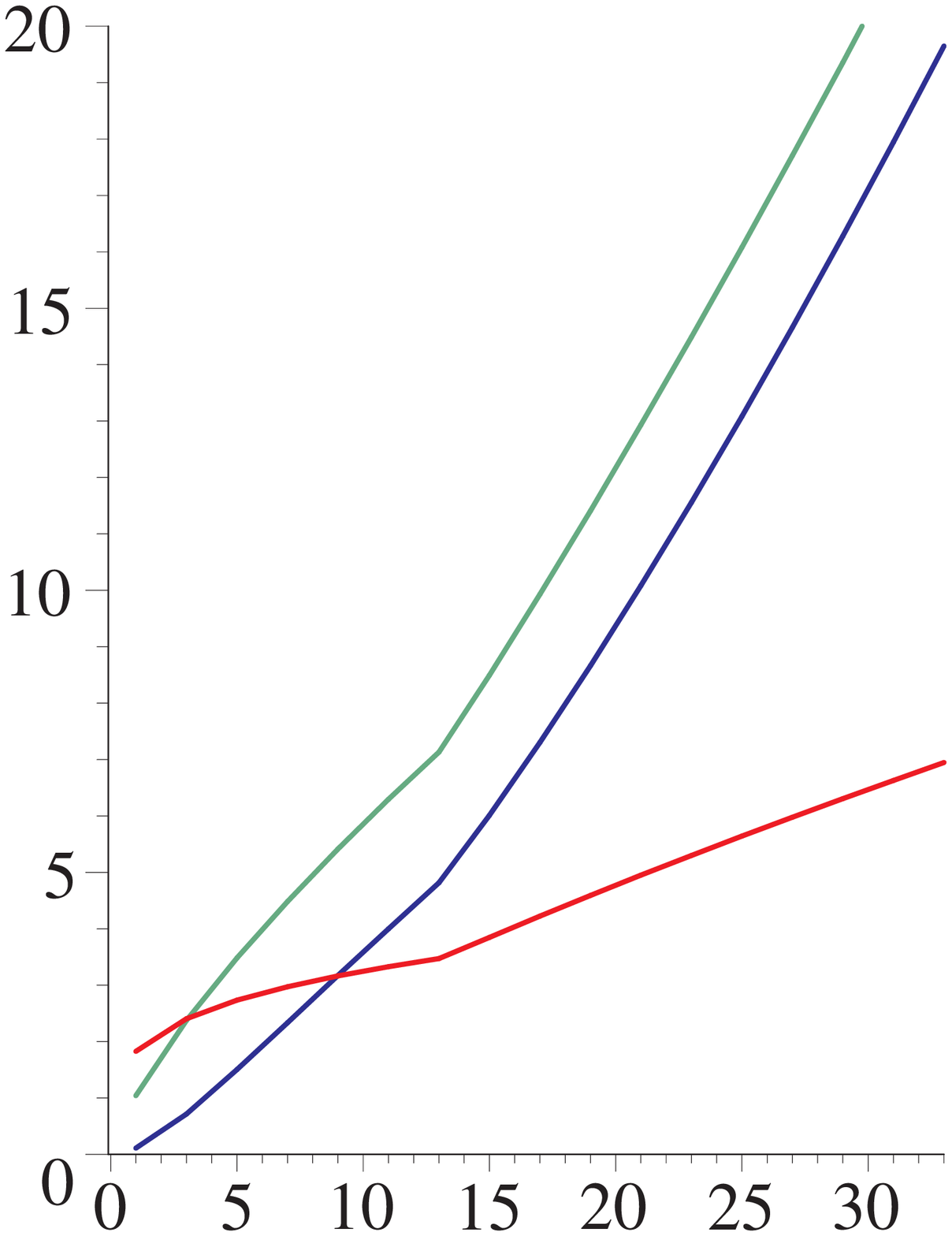}
     \put(-137,115){\rotatebox{0}{\mbox{$[\lambda_L]$}}}
     \put(-30,-10){$2J+1$}
  \end{minipage}
\end{center}
\caption{(Color online) The left panel illustrates the behavior of
the relative
  deviation of the intensity distribution $\Delta \bar I$ from zero as
  it approaches the 0.74\%-deviation marks (top and bottom
  grid). Here, $\Delta \bar I_{23}$ is shown for the crossed
  configuration of two laser beams travelling along $z$ and $x$-axis
  respectively. The value of the Rayleigh length $z_{Rx}$ at which we
  find that the oscillatory behaviour of $\Delta \bar I$ along a
  constant radial perimeter just exhausts the upper and lower limits
  set by the deviation marks allows us to determine the associated
  value of $z_{\min}$. The latter is plotted as a function of maximum
  mode number, in the middle panel (the filled in red area is the
  forbidden area of too tightly focussed beams). The values of
  $z_{\min}(2J+1)$ in turn determine the
  positions of the turning points $0.57\cdot
  \sqrt{2J+1}\cdot w_{0x}$ (top green line), the positions of the
  deviation-points~$d_{2J+1}$ (middle blue line)
  and the minimal beam widths $w_{0x}(2J+1)$ (bottom red line),
  depicted in the right panel in units of the laser's wavelength
  $\lambda_L$.}
\label{Fig_Iradial_zMin_alphaMax}
\end{figure}
Gouy's phase~$\phi(z)=\arctan(z/z_R)\approx z /z_R,$ introduces
relative phases between the modes within each beam. Since the
Gouy-phase varies strongest near the beam focus we have to consider
its mode dispersive effects~\cite{Ole_2005AmJPh}. If the beam is
very strongly focussed (small value of~$z_{R_x}$) the dephasing away
from the focus $z=0$ is so rapid that non-linear aberrations degrade
the desired linear field profile over the width of the atom-lens. In
other words, a lower limit for the Rayleigh lengths~$z_{\min}(2J+1)$
as a function of the number of used modes~`$2J+1$' has to be
determined in order to guarantee moderate dephasing. Whereas the
absolute values for this lower limit are hard to derive from first
principles, we can still work out the correct scaling with the mode
number:

The electric field is proportional to the superposition of the modes
including the Gouy-phase factors; this can be approximated by
$E_{2J+1} \propto \sum_{j=1,3,...}^{2J+1} c_j \psi_j e^{i j \phi}
\approx \sum_{j=1,3,...}^{2J+1} c_j \psi_j ( 1 + i j z/z_{R_x}) $.
The expansion coefficients are positive and the wave functions are
real at the focus $z=0$. Since the first order term is purely
imaginary the integrated intensity has to depend on $z$
quadratically: $\bar I_{2J+1}(z) = \bar I_{2J+1}(0) \cdot
[1+\frac{z^2}{z^2_{R_x}} F_{2J+1} + {\cal O}(z^4)]$. The term
$F_{2J+1}$ has a complicated dependence on the number of modes, but,
containing the square of sums of the form~$\sum_{j=1,3,...}^{2J+1} j
c_j \psi_j $, is roughly proportional to $(2J+1)^2$. When we
consider the relative deviation of the intensity profile near the
focus from the focal intensity distribution, $\Delta \bar I =
\frac{\bar I(z)-\bar I(0)}{\bar I(0)}$, we find $\Delta \bar
I_{2J+1} \propto \frac{z^2}{z^2_{R_x}} \cdot (2J+1)^2 $.
Additionally, we know that the widths of the superpositions scale
roughly like those of the harmonic oscillator~\cite{Ole_2005AmJPh},
see Fig.~\ref{fig_I_TurningPoints}, namely $z\propto \sqrt{2J+1}$.
For constant relative intensity deviations $\Delta \bar I_{2J+1}$
this implies $const. = \frac{\sqrt{2J+1}^2}{z^2_{R_x}} \cdot
(2J+1)^2$ or $z_{R_x}\propto (2J+1)^{3/2}$. A numerical
investigation, see Fig.~\ref{Fig_Iradial_zMin_alphaMax}, confirms
$z_{\min}(2J+1) = 0.8 \cdot \lambda_L \cdot (2J+1)^{3/2}$ as a good
estimate for a lower bound on~$z_{R_x}$. This relationship has been
checked numerically and holds for $15 < 2 J + 1 < 55$. There is no
reason to believe deviations might occur for values of $ 2J +1 >
55$, but for small values of $J$ the assumptions used in the
derivation of the scaling law do not hold accurately, see
Fig.~\ref{fig_I_TurningPoints}. Instead, the expression
$z_{\min}(2J+1) = 10.5 \cdot \lambda_L \cdot (2J+1)^{1/2}$ gives a
much better estimate for $z_{\min}(2J+1)$ in the range of $1 < 2 J +
1 \leq 13$. These lower limits for $z_{R_x}$ imply that the beam focus
is several wavelengths wide and {a posteriori} confirms that the
paraxial approximations hold for all cases discussed here, since the
largest beam opening angle conforming with the lower limits
presented here turns out to be roughly $7.5^\circ$ for
superposition~$\Psi_3$.

\IgnoreThis{\section{Conclusions}\label{S_Conclusion}}

For a possible experimental implementation of the ideas presented
here it should be emphasized that throughout the use of a repulsive
(blue-detuned) optical potential has been assumed since it allows us
to build focussing lenses with a dark center reducing detrimental
spontaneous emission noise.  Equivalent logic applies to `concave'
atomic lenses which would best be implemented in red detuning, with
dark centers as well.

The Raman-Nath assumption becomes progressively worse the larger the
numerical aperture of a lens.  Trajectory simulations show a
`downhill' drift of atomic paths that can partly be compensated for by
slightly weakening the rise of the potential through the suitable
subtraction of higher-order terms that lead to slight non-harmonic
modifications of the lens, improving its performance. Clearly, if such
fine-tuning is considered, the approximations
underlying~Eq.~(\ref{Intensity}) and
Eq.~(\ref{dipole_potential_approx}) might not be permissible. These
considerations are beyond the scope of this paper.

The techniques for the coherent superposition of laser modes have
been experimentally demonstrated, see e.g.
references~\cite{OleJMO05,Maurer_RitschMarte_NJP07} and citations
therein. We have found here that using the mode-superposition
approach allows for very considerable laser power savings and lenses
can be made wider than is possible with pure modes. We come to the
conclusion that for the design of atomic lenses, based on the
optical dipole force, it is possible and necessary to coherently
superpose suitable laser modes in order to create wide thin
parabolic lenses with large numerical apertures.

Obviously the approach presented here can be applied for the
manipulation of stationary atomic clouds just as well as for atomic
beams~\cite{Ole_JOPA05,Murray_JPhB05}.

\IgnoreThis{\section*{Acknowledgments}}

I would like to thank Neil Oxtoby for careful reading of the
manuscript and the referees for their very helpful comments.
%

\bibliography{OpticalAtomLensPaperBib}

\end{document}